\documentclass[a4paper,11pt]{article}

\usepackage{jheppub} 
\usepackage{subfigure}
\usepackage{hyperref}
\usepackage[T1]{fontenc} 

\title{\boldmath localization mechanism of the Kalb-Ramond field on brane with codimension-two }

\author[a]{Yong-Tao Lu,}
\author[a,1]{Heng Guo,\note{Corresponding author.}}
\author[a]{Qun Wei,}
\author[a,2]{and Bing Wei.\note{Corresponding author.}}
\affiliation[a]{School of Physics,
              Xidian University,
              Xi'an 710071, People's Republic of China}
\emailAdd{luyt@stu.xidian.edu.cn}
\emailAdd{hguo@xidian.edu.cn}
\emailAdd{qunwei@xidian.edu.cn}
\emailAdd{bwei@xidian.edu.cn}

\abstract{
The $2$-form Kalb-Ramond (KR) field, together with the metric tensor and dilaton, arises as one of the massless
excitation mode of a closed string. Subsequently, this field plays an important role in both string theory and
field theory. In this paper, we investigate the localization of the KR field  on the brane with codimension-2.
A general Kaluza-Klein (KK) decomposition is adopted, wherein the six-dimensional KR field is expanded into
one four-dimensional (4D) KR field, two 4D vector fields, and one 4D scalar field. Then, for the case of the
extra dimensions $\mathcal{R}_1\times\mathcal{R}_1$, only the 4D scalar field can be localized on the brane.
In contrast, for the case of extra dimensions $\mathcal{R}_1\times\mathcal{S}_1$, one 4D vector field
and the 4D scalar field can be localized on the brane at the same time. In both cases, the mass of the 4D 
scalar field remains zero. Next, we examine the localization of
the KR field within a specific six-dimensional brane model with extra dimensions $\mathcal{R}_1\times\mathcal{S}_1$.
By introducing the background scalar coupling, we show that the 4D KR field, along with the other three
4D fields, can be localized on the brane under the condition of the coupling parameter $t>v^2/12$. Additionally
in this case, for both the 4D KR field and the one 4D vector field which acquires its mass from the non-compact
extra dimension, the resonant KK modes could exist near the origin of this extra dimension.
}

\begin{document}
\maketitle
\flushbottom

\section{Introduction}

As an extra-dimensional theory, the braneworld scenario offers an alternative framework to address several
puzzling longstanding puzzles, such as the hierarchy problem \cite{PLB429263,PLB436257,PRLRS83,
NJP201012,PRD201184,PRD201387} and the cosmological constant problem \cite{PLB1983125,PLB125139,
PLB1986166,NPB2000584,PLB2000480,PRD200062021,PLB2000474,PRL200186,PRL200187,PLB2004600}. In the Randall-Sundrum
(RS) braneworld model \cite{PRL199983RS}, effective four-dimensional (4D) gravity can be recovered even in the
presence of an infinite extra dimension. However, this model involves singularities at the brane position, which
arise from the idealization of the brane as being infinitely thin. To address this issue, smooth thick braneworld
solutions have been extensively studied, typically involving gravity coupled to background scalar
\cite{PRL1999834922,PLB2000478,PRD200062008,NPB2000581,PRL200288141602,PRD0265014,PRD0266024,PRD0367003,
JHEP0405012,PLB06634526,PRD1081008,JHEP1106135,PRD1285033,PRD1491016,PRD1489004,WJJjhep2105017,NakasPRD109}.
In addition, there are also thick branes derived from pure geometry \cite{JHEP0510101,JHEP1010069,JHEP1004130,
JHEP1202083,PRDYJ1285033,EPJC1676321,PRDGH23107017,PLBGH2410.11310}, as well as from couplings with other types
of matter fields \cite{PRDGWJ1693035,EPJCCZQ2383275,EPJCDV238355}.

It is known that the Kalb-Ramond (KR) field (NS-NS B-field) is an antisymmetric tensor field with higher
spin proposed in string theory. This field appears as a massless excitation mode in the
low-energy limit of closed strings, and can describe axion physics \cite{WittenPLB1984,Witten0605206} or the spacetime torsion
of a Riemannian manifold \cite{PRLBM0289101,PRDBM0470009}.
The rank-2 antisymmetric KR field serves as the gauge field of strings and, in heterotic string theory, coupled to the
Yang-Mills-Chern-Simons 3-form \cite{KalbPRD1074,G-S-W}.
In 4D spacetime, the KR field is dual to the
scalar field, while in higher-dimensional spacetimes, it represents additional particles.

In braneworld scenarios, a significant and intriguing area of exploration involves the study of field
localization \cite{PRD201387,PRD1285033,WJJjhep2105017,PRDGH23107017,PLBGY00474,PLBBG00474,PLBSR00492,
PLBO00496,PRDLYX0904.1785,PRDLYX0907.0910,JHEPLYX1002080,JHEPFCE1210060,PRDGA1693064,ZZHjhep1805072,
FCEjhep1901021,JHEPGH2406,PRDLYT25111025}. In refs. \cite{PRLBM0289101,PRDBM0470009}, the KR field is
considered as a representation of bulk torsion and is found to have a significantly weaker effect
than curvature. This property makes its detection particularly challenging on the RS thin brane. However,
when treated as a type of matter field, the KR field can be localized in certain cases
of five-dimensional brane models. For the Minkowski brane case, the KR field can be localized on the
thick brane through dilaton coupling \cite{PRDLYX1285023,PRDMOT0979085,ELWTC0988001,PRDFCE1184036,
EPJCWTC137323} or background scalar coupling \cite{PRDDYZ1388009}. Reference \cite{PRDLYT25111025}
further shows that introducing the coupling between the KR field and the gravity enables its localization
on thick brane in both cases of Minkowski and Anti-de Sitter brane. For the de Sitter brane case,
localization of the KR field can be realized via coupling with the background scalar \cite{CTPYC2072801}.
For comprehensive reviews, see refs. \cite{ExtraDimensionLiu,PRDVA22967}. Notably, most of the aforementioned
works focus exclusively on codimension-1 scenarios. Regarding the higher-codimension case, refs. \cite{WJJjhep2105017,FCEjhep1901021,arXiv1912.03859,PLBfce20810781,MWXarXiv2024,CPCfuce2510.1088,
JHEPWJJ2312033} explored the localization of various matter fields in bulk with more than one codimension.
In refs. \cite{WJJjhep2105017,FCEjhep1901021,JHEPWJJ2312033}, the scalar field, the vector field and
the spinor field are studied in the six-dimensional (6D) bulk spacetime with codimension-2. However,
studies focusing on the KR field in such higher-codimension cases remain scarce.

In this paper, we investigate the localization of the KR field on the codimension-2 brane in two cases
of the extra-dimensional manifold: $\mathcal{R}_1\times\mathcal{R}_1$ and $\mathcal{R}_1\times\mathcal{S}_1$.
A general KK decomposition is employed, under which the 6D KR field decomposes into four 4D components:
one 4D KR field, two 4D vector fields, and one 4D scalar field. It is demonstrated that the mass of the
4D KR field comes from both two extra dimensions, while the two 4D vector fields acquire masses independently
from the respective single extra dimension, and the 4D scalar field remains massless. Besides, there are
five types of interactions between these four 4D fields. Then, we observe that in the minimal coupling case,
the localization of these four 4D fields is independent from the concrete brane model. Specifically, for
the case of extra dimensions $\mathcal{R}_1\times\mathcal{R}_1$, only the 4D scalar field can be localized
on the brane. In contrast, for the $\mathcal{R}_1\times\mathcal{S}_1$ case, one 4D vector field whose mass
originates from the compact extra dimension, and the 4D scalar field, can be localized. Lastly, we consider
a concrete 6D thick brane model incorporating the background scalar coupling. Under the condition of the 
coupling parameter $t>v^2/12$, all 4D components-the KR field, the two vector fields, and the scalar 
field-can be localized on the brane. For both the 4D KR field and the one 4D vector field whose mass arises 
from the non-compact extra dimension, resonant KK modes could exist, with their numbers increasing as the 
parameter $t$ increases.

The paper is structured as follows: In section \ref{method}, we perform the KK decomposition of the 6D KR
field and derive the effective 4D action. The case of the extra-dimensional manifold
$\mathcal{R}_1\times\mathcal{R}_1$ is discussed detailedly in section \ref{sec2-1}, while the case of
the extra-dimensional manifold $\mathcal{R}_1\times\mathcal{S}_1$ is analyzed in section \ref{sec2-2}.
In section \ref{Loc}, we investigate the localization of the KR field within a concrete 6D brane model.
Lastly, the conclusions are summarized in section \ref{Cons}.


\section{KK decomposition and the effective action }\label{method}

For a 6D massless KR field $B_{MN}$, its action is assumed as
\begin{eqnarray} \label{action0}
  S_{\text{KR}}=-\int d^6x\sqrt{-g}H^{MNL}H_{MNL},
\end{eqnarray}
where the field strength
\begin{eqnarray} \label{fieldstrength}
  H_{MNL}=\partial_M B_{NL}+\partial_L B_{MN}+\partial_N B_{LM}.
\end{eqnarray}
The capital Latin letters $M,N,L=0,1,2,3,5,6$ are used to represent the bulk indices.

For the localization of this KR field, we will consider two cases of the extra dimensions. The first
case is the extra dimensions $\mathcal{R}_1\times\mathcal{R}_1$, and the other one is
$\mathcal{R}_1\times\mathcal{S}_1$. The latter case includes a compact extra dimension with enough
small radius. For both two cases, we will choose the general KK decomposition with no gauge choice.

\subsection{The extra dimensions $\mathcal{R}_1\times\mathcal{R}_1$} \label{sec2-1}

In this subsection, we will discuss the localization of the KR field in the case of two infinite extra dimensions.
The line element of the 6D bulk is assumed as
\begin{eqnarray} \label{metricyz}
  ds^2=a^2(y,z)(\hat g_{\mu\nu}dx^{\mu}dx^{\nu}+dy^2+dz^2),
\end{eqnarray}
where $y$ and $z$ represent the two infinite extra dimensions, $a(y,z)$ is the warp factor, and $\hat g_{\mu\nu}$
is the metric on the brane with $\mu,\nu=0,1,2,3$.

Concerning the 6D KR field $B_{MN}$, we perform the following general KK decomposition
\begin{subequations}\label{decomposition1yz}
\begin{eqnarray}
      B_{\mu\nu}&=&\sum_m \hat B_{\mu\nu}^{(m)}(x^{\sigma})W_1^{(m)}(y,z),              \\
      B_{\mu y} &=&\sum_m \hat C_{\mu}^{(m)}(x^{\sigma})W_2^{(m)}(y,z),                 \\
      B_{\mu z} &=&\sum_m \hat D_{\mu}^{(m)}(x^{\sigma})W_3^{(m)}(y,z),                 \\
      B_{y z}   &=&\sum_m \hat{\zeta}^{(m)}(x^{\sigma})W_4^{(m)}(y,z).
 \end{eqnarray}
\end{subequations}
From these expressions, it can be seen that $\hat B_{\mu\nu}^{(m)}(x^{\sigma})$ is the 4D KR field,
$\hat C_{\mu}^{(m)}(x^{\sigma})$ and $\hat D_{\mu}^{(m)}(x^{\sigma})$ are two 4D vector fields, and
$\hat{\zeta}^{(m)}(x^{\sigma})$ is the 4D scalar field. For this general KK decomposition (\ref{decomposition1yz}), no gauge choice has been
introduced, so the factors $W_2^{(m)}(y,z)$, $W_3^{(m)}(y,z)$, and $W_4^{(m)}(y,z)$ do not always vanish,
and the two 4D vector fields and the 4D scalar field make differences on the brane. Besides, under
a similar KK decomposition, ref. \cite{arXiv1912.03859} has demonstrated the gauge invariance of the effective action
for a massless $q-$form field on a $p-$brane with codimension-two.

Based on the KK decomposition (\ref{decomposition1yz}), the 6D action (\ref{action0}) can be reduced to
the following 4D effective one
\begin{eqnarray}\label{4Dactionyz}
 S_{\text{KR}}&=& -\int d^6x\sqrt{-g}H^{MNL}H_{MNL}                                              \nonumber   \\
              &=& -\int d^6x\sqrt{-g}\big(H^{\mu\nu\tau}H_{\mu\nu\tau}+2H^{\mu\nu y}H_{\mu\nu y}
                   +2H^{\mu\nu z}H_{\mu\nu z}+2H^{\mu yz}H_{\mu yz}\big)           \nonumber   \\
              &=& -\sum_m\sum_{m'}\int d^4x\sqrt{-\hat g}\bigg[I_1^{(mm')}\hat H^{\mu\nu\tau(m)}\hat H_{\mu\nu\tau}^{(m')}
                   +\big(I_5^{(mm')}+I_9^{(mm')}\big)\hat B^{\mu\nu(m)}\hat B_{\mu\nu}^{(m')}        \nonumber   \\
              & & +I_2^{(mm')}\hat F^{\mu\nu(m)}\hat F_{\mu\nu}^{(m')}+I_6^{(mm')}\hat G^{\mu\nu(m)}\hat G_{\mu\nu}^{(m')}
                   +I_3^{(mm')}\hat B^{\mu\nu(m)}\hat F_{\mu\nu}^{(m')}                              \nonumber   \\  
              & & +I_4^{(mm')}\hat F^{\mu\nu(m)}\hat B_{\mu\nu}^{(m')}+I_7^{(mm')}\hat B^{\mu\nu(m)}\hat G_{\mu\nu}^{(m')}
                   +I_8^{(mm')}\hat G^{\mu\nu(m)}\hat B_{\mu\nu}^{(m')}                              \nonumber   \\  
              & & +I_{10}^{(mm')}\partial^{\mu}\hat{\zeta}^{(m)}\partial_{\mu}\hat{\zeta}^{(m')}
                   +I_{11}^{(mm')}\hat C^{\mu(m)}\partial_{\mu}\hat{\zeta}^{(m')}
                   -I_{12}^{(mm')}\hat D^{\mu(m)}\partial_{\mu}\hat{\zeta}^{(m')}                    \nonumber   \\  
              & & +I_{13}^{(mm')}\partial^{\mu}\hat{\zeta}^{(m)}\hat C_{\mu}^{(m')}
                   +I_{14}^{(mm')}\hat C^{\mu(m)}\hat C_{\mu}^{(m')}
                   -I_{15}^{(mm')}\hat D^{\mu(m)}\hat C_{\mu}^{(m')}                                 \nonumber   \\  
              & & -I_{16}^{(mm')}\partial^{\mu}\hat{\zeta}^{(m)}\hat D_{\mu}^{(m')}
                   -I_{17}^{(mm')}\hat C^{\mu(m)}\hat D_{\mu}^{(m')}
                   +I_{18}^{(mm')}\hat D^{\mu(m)}\hat D_{\mu}^{(m')}\bigg],
\end{eqnarray}
where the field strengths
\begin{subequations}\label{fieldstrengthyz}
\begin{eqnarray}
    \hat H_{\mu\nu\lambda}^{(m)}&=&\partial_{\mu}\hat B_{\nu\lambda}^{(m)}
           +\partial_{\lambda}\hat B_{\mu\nu}^{(m)}+\partial_{\nu}\hat B_{\lambda\mu}^{(m)},      \\
    \hat F_{\mu\nu}^{(m)}&=&\partial_{\mu}\hat C_{\nu}^{(m)}
           -\partial_{\nu}\hat C_{\mu}^{(m)},                 \\
    \hat G_{\mu\nu}^{(m)}&=&\partial_{\mu}\hat D_{\nu}^{(m)}
           -\partial_{\nu}\hat D_{\mu}^{(m)},
 \end{eqnarray}
\end{subequations}
and the matrices
\begin{subequations} \label{integral}
\begin{eqnarray}
  I_1^{(mm')}&\equiv& \int d^2xW_1^{(m)}W_1^{(m')},                                               \\
  I_2^{(mm')}&\equiv& 2\int d^2xW_2^{(m)}W_2^{(m')},                                              \\
  I_3^{(mm')}&\equiv& 2\int d^2x\left(\partial_yW_1^{(m)}\right)W_2^{(m')},                       \\
  I_4^{(mm')}&\equiv& 2\int d^2xW_2^{(m)}\left(\partial_yW_1^{(m')}\right),                       \\
  I_5^{(mm')}&\equiv& 2\int d^2x\left(\partial_yW_1^{(m)}\right)\left(\partial_yW_1^{(m')}\right),             \\
  I_6^{(mm')}&\equiv& 2\int d^2xW_3^{(m)}W_3^{(m')},                                              \\
  I_7^{(mm')}&\equiv& 2\int d^2x\left(\partial_zW_1^{(m)}\right)W_3^{(m')},                       \\
  I_8^{(mm')}&\equiv& 2\int d^2xW_3^{(m)}\left(\partial_zW_1^{(m')}\right),                       \\
  I_9^{(mm')}&\equiv& 2\int d^2x\left(\partial_zW_1^{(m)}\right)\left(\partial_zW_1^{(m')}\right),             \\
  I_{10}^{(mm')}&\equiv& 2\int d^2xW_4^{(m)}W_4^{(m')},                                           \\
  I_{11}^{(mm')}&\equiv& 2\int d^2x\left(\partial_zW_2^{(m)}\right)W_4^{(m')},                    \\
  I_{12}^{(mm')}&\equiv& 2\int d^2x\left(\partial_yW_3^{(m)}\right)W_4^{(m')},                    \\
  I_{13}^{(mm')}&\equiv& 2\int d^2xW_4^{(m)}\left(\partial_zW_2^{(m')}\right),                    \\
  I_{14}^{(mm')}&\equiv& 2\int d^2x\left(\partial_zW_2^{(m)}\right)\left(\partial_zW_2^{(m')}\right),          \\
  I_{15}^{(mm')}&\equiv& 2\int d^2x\left(\partial_yW_3^{(m)}\right)\left(\partial_zW_2^{(m')}\right),          \\
  I_{16}^{(mm')}&\equiv& 2\int d^2xW_4^{(m)}\left(\partial_yW_3^{(m')}\right),                    \\
  I_{17}^{(mm')}&\equiv& 2\int d^2x\left(\partial_zW_2^{(m)}\right)\left(\partial_yW_3^{(m')}\right),          \\
  I_{18}^{(mm')}&\equiv& 2\int d^2x\left(\partial_yW_3^{(m)}\right)\left(\partial_yW_3^{(m')}\right).
\end{eqnarray}
\end{subequations}
The self-interaction between the same particles leads to the mass. For this effective action (\ref{4Dactionyz}),
the factor $(I_5^{(mm')}+I_9^{(mm')})$ corresponds to the mass of the 4D KR field $\hat B_{\mu\nu}$, and
factors $I_{14}^{(mm')},I_{18}^{(mm')}$ correspond to the masses of the 4D vector fields $\hat C_{\mu}$ and
$\hat D_{\mu}$, respectively. For the 4D scalar field $\hat{\zeta}$, its mass always be zero, and there is
no self-interaction.

Furthermore, the other factors describe the interactions between these 4D fields:
\begin{itemize}
  \item $I_3^{(mm')},I_4^{(mm')}$ denote the interaction between the KR field $\hat B_{\mu\nu}$ and the vector
          field $\hat C_{\mu}$;
  \item $I_7^{(mm')},I_8^{(mm')}$ denote the interaction between the KR field $\hat B_{\mu\nu}$ and the vector
          field $\hat D_{\mu}$;
  \item $I_{11}^{(mm')},I_{13}^{(mm')}$ denote the interaction between the vector field $\hat C_{\mu}$ and the
          scalar field $\hat{\zeta}$;
  \item $I_{12}^{(mm')},I_{16}^{(mm')}$ denote the interaction between the vector field $\hat D_{\mu}$ and the
          scalar field $\hat{\zeta}$;
  \item $I_{15}^{(mm')},I_{17}^{(mm')}$ denote the interaction between the two vector fields $\hat C_{\mu}$
          and $\hat D_{\mu}$.
\end{itemize}
In terms of eqs. (\ref{integral}), it can be seen that for the above five items, the two matrices within
each one are mutually transposed. Additionally, we can see that there is no interaction between the 4D KR
field $\hat B_{\mu\nu}$ and the 4D scalar field $\hat{\zeta}$.

Moving forward, varying this effective action (\ref{4Dactionyz}) with respect to the 4D fields $\hat B_{\mu\nu}^{(m)}$,
$\hat C_{\mu}^{(m)}$,$\hat D_{\mu}^{(m)}$, and $\hat{\zeta}^{(m)}$, we can get
\begin{subequations}  \label{varybranefield}
\begin{align}
  \frac{I_1^{(mm')}}{\sqrt{-\hat g}}\partial_{\lambda}\hspace{-0.02cm}\left(\sqrt{-\hat g}\hat H^{\mu\nu\lambda(m)}\right) \hspace{-0.05cm}
     -\hspace{-0.05cm}\left(I_5^{(mm')}\hspace{-0.05cm}+I_9^{(mm')}\right)\hat B^{\mu\nu(m)}\hspace{-0.05cm}
     -I_4^{(mm')}\hat F^{\mu\nu(m)}\hspace{-0.05cm}                         
     -\hspace{-0.05cm}I_8^{(mm')}\hat G^{\mu\nu(m)}=0,                                                                       \\
  \frac{I_2^{(mm')}}{\sqrt{-\hat g}}\partial_{\nu}\left(\sqrt{-\hat g}\hat F^{\mu\nu(m)}\right)
     +\frac{I_3^{(mm')}}{\sqrt{-\hat g}}\partial_{\nu}\left(\sqrt{-\hat g}\hat B^{\mu\nu(m)}\right)
     +I_{13}^{(mm')}\partial^{\mu}\hat{\zeta}^{(m)}+I_{14}^{(mm')}\hat C_{\mu}^{(m)}         \hspace{1cm}         \nonumber   \\
     -I_{15}^{(mm')}\hat D_{\mu}^{(m)}=0,                                   \\
  \frac{I_6^{(mm')}}{\sqrt{-\hat g}}\partial_{\nu}\left(\sqrt{-\hat g}\hat G^{\mu\nu(m)}\right)
     +\frac{I_7^{(mm')}}{\sqrt{-\hat g}}\partial_{\nu}\left(\sqrt{-\hat g}\hat B^{\mu\nu(m)}\right)
     -I_{16}^{(mm')}\partial^{\mu}\hat{\zeta}^{(m)}-I_{17}^{(mm')}\hat C_{\mu}^{(m)}         \hspace{1cm}         \nonumber    \\
     +I_{18}^{(mm')}\hat D_{\mu}^{(m)}=0,                                   \\
  I_{10}^{(mm')}\partial_{\mu}\left(\sqrt{-\hat g}\partial^{\mu}\hat{\zeta}^{(m)}\right)
     +I_{11}^{(mm')}\partial_{\mu}\left(\sqrt{-\hat g}\hat C_{\mu}^{(m)}\right)
     -I_{12}^{(mm')}\partial_{\mu}\left(\sqrt{-\hat g}\hat D_{\mu}^{(m)}\right)=0.
\end{align}
\end{subequations}

On the other hand, we vary the 6D action (\ref{action0}) with respect to the 6D KR field $B_{MN}$, obtaining
the equation of motion
\begin{eqnarray} \label{EoM}
  \frac{1}{\sqrt{-g}}\partial_M\left(\sqrt{-g}H^{MNL}\right)=0.
\end{eqnarray}
This equation of motion can be expanded into the following component equations
\begin{subequations} \label{componentEoM}
\begin{align}
  &\frac{1}{\sqrt{-\hat g}}\partial_{\lambda}\left(\sqrt{-\hat g}\hat H^{\mu\nu\lambda(m)}\right)
    +(\lambda_1+\lambda_2)\hat B^{\mu\nu(m)}+\lambda_3\hat F^{\mu\nu(m)}+\lambda_4\hat G^{\mu\nu(m)}=0,        \\
  &\frac{1}{\sqrt{-\hat g}}\partial_{\nu}\left(\sqrt{-\hat g}\hat F^{\mu\nu(m)}\right)
    +\frac{\lambda_5}{\sqrt{-\hat g}}\partial_{\nu}\left(\sqrt{-\hat g}\hat B^{\mu\nu(m)}\right)
    -\lambda_6\partial^{\mu}\hat{\zeta}^{(m)}-\lambda_7\hat C^{\mu(m)}+\lambda_8\hat D^{\mu(m)}=0,             \\
  &\frac{1}{\sqrt{-\hat g}}\partial_{\nu}\left(\sqrt{-\hat g}\hat G^{\mu\nu(m)}\right)
    +\frac{\lambda_9}{\sqrt{-\hat g}}\partial_{\nu}\left(\sqrt{-\hat g}\hat B^{\mu\nu(m)}\right)  \hspace{-0.05cm}
    +\lambda_{10}\partial^{\mu}\hat{\zeta}^{(m)}+\lambda_{11}\hat C^{\mu(m)}                     \hspace{-0.05cm}
    -\lambda_{12}\hat D^{\mu(m)}=0,                                                                            \\
  &\partial_{\mu}\left(\sqrt{-g}\partial^{\mu}\hat{\zeta}^{(m)}\right)
    +\lambda_{13}\partial_{\mu}\left(\sqrt{-g}\hat C^{\mu(m)}\right)
    -\lambda_{14}\partial_{\mu}\left(\sqrt{-g}\hat D^{\mu(m)}\right)=0,
\end{align}
\end{subequations}
where
\begin{eqnarray} \label{lambda}
  &&\lambda_1 \equiv\frac{\partial^2_yW_1^{(m)}}{W_1^{(m)}},  \hspace{1.4cm}
    \lambda_2 \equiv\frac{\partial^2_zW_1^{(m)}}{W_1^{(m)}},  \hspace{1.4cm}
    \lambda_3 \equiv\frac{\partial_yW_2^{(m)}}{W_1^{(m)}},    \hspace{1.4cm}
    \lambda_4 \equiv\frac{\partial_zW_3^{(m)}}{W_1^{(m)}},                       \nonumber      \\
  &&\lambda_5 \equiv\frac{\partial_yW_1^{(m)}}{W_2^{(m)}},    \hspace{1.4cm}
    \lambda_6 \equiv\frac{\partial_zW_4^{(m)}}{W_1^{(m)}},    \hspace{1.4cm}
    \lambda_7 \equiv\frac{\partial^2_zW_2^{(m)}}{W_2^{(m)}},  \hspace{1.4cm}
    \lambda_8 \equiv\frac{\partial_z\partial_yW_3^{(m)}}{W_2^{(m)}},             \nonumber      \\
  &&\lambda_9 \equiv\frac{\partial_zW_1^{(m)}}{W_3^{(m)}},      \hspace{1.35cm}
    \lambda_{10}\equiv\frac{\partial_yW_4^{(m)}}{W_3^{(m)}},    \hspace{1.23cm}
    \lambda_{11}\equiv\frac{\partial_y\partial_zW_2^{(m)}}{W_3^{(m)}},  \hspace{0.94cm}
    \lambda_{12}\equiv\frac{\partial^2_yW_3^{(m)}}{W_3^{(m)}},                  \nonumber      \\
  &&\lambda_{13}\equiv\frac{\partial_zW_2^{(m)}}{W_4^{(m)}},    \hspace{1.2cm}
    \lambda_{14}\equiv\frac{\partial_yW_3^{(m)}}{W_4^{(m)}}.
\end{eqnarray}
Here, $\lambda_i(i=1,2,\cdots, 14)$ contain the order $m$, and are not required to be constants.

At this stage, since both equations (\ref{varybranefield}) and (\ref{componentEoM}) are derived from the 6D action
(\ref{action0}), these two sets of equations should be compatible with each other. Therefore, there are
consistency conditions:
\begin{eqnarray} \label{lambdaI}
  &&I_1^{(mm')}=\delta^{mm'},                    \hspace{1.2cm}
    I_2^{(mm')}=\delta^{mm'},                    \hspace{1.2cm}
    I_3^{(mm')}=\lambda_5\delta^{mm'},           \hspace{0.8cm}
    I_4^{(mm')}=-\lambda_3\delta^{mm'},                              \nonumber    \\
  &&I_5^{(mm')}=-\lambda_1\delta^{mm'},          \hspace{0.5cm}
    I_6^{(mm')}=\delta^{mm'},                    \hspace{1.2cm}
    I_7^{(mm')}=\lambda_9\delta^{mm'},           \hspace{0.8cm}
    I_8^{(mm')}=-\lambda_4\delta^{mm'},                              \nonumber    \\
  &&I_9^{(mm')}=-\lambda_2\delta^{mm'},          \hspace{0.5cm}
    I_{10}^{(mm')}=\delta^{mm'},                 \hspace{1.2cm}
    I_{11}^{(mm')}=\lambda_{13}\delta^{mm'},     \hspace{0.6cm}
    I_{12}^{(mm')}=\lambda_{14}\delta^{mm'},                         \nonumber    \\
  &&I_{13}^{(mm')}=-\lambda_{6}\delta^{mm'},     \hspace{0.5cm}
    I_{14}^{(mm')}=-\lambda_{7}\delta^{mm'},     \hspace{0.5cm}
    I_{15}^{(mm')}=-\lambda_{8}\delta^{mm'},     \hspace{0.5cm}
    I_{16}^{(mm')}=-\lambda_{10}\delta^{mm'},                        \nonumber    \\
  &&I_{17}^{(mm')}=-\lambda_{11}\delta^{mm'},    \hspace{0.39cm}
    I_{18}^{(mm')}=-\lambda_{12}\delta^{mm'}.
\end{eqnarray}
As mentioned above, there are five sets of transposition matrices in eqs. (\ref{integral}), which describe
the interactions between the four 4D fields. Subsequently, under the conditions (\ref{lambdaI}), the
two matrices within each set are equal, and the same for the corresponding two $\lambda_i$, reads
\begin{eqnarray} \label{lambdaIEq}
  & &\lambda_3=\lambda_5,\ \ \
     \lambda_4 =\lambda_9,\ \ \
     \lambda_{10}=\lambda_{14},\ \ \                \nonumber     \\
  & &\lambda_6=\lambda_{13},\ \ \
     \lambda_8=\lambda_{11}.
\end{eqnarray}

Then, from eqs. (\ref{lambda}), we can get
\begin{subequations} \label{Schroyz}
\begin{eqnarray}
  \partial_y^2W_1^{(m)}&=\lambda_1W_1^{(m)},                                     \\
  \partial_z^2W_1^{(m)}&=\lambda_2W_1^{(m)},                                     \\
  \partial_z^2W_2^{(m)}&=\lambda_7W_2^{(m)},                                     \\
  \partial_y^2W_3^{(m)}&=\lambda_{12}W_3^{(m)},
\end{eqnarray}
\end{subequations}
and
\begin{subequations}\label{1stEqtyz}
\begin{align}
  \partial_yW_2^{(m)}&=\lambda_3W_1^{(m)},                    \\
  \partial_zW_3^{(m)}&=\lambda_4W_1^{(m)},                    \\
  \partial_yW_1^{(m)}&=\lambda_5W_2^{(m)},                    \\
  \partial_zW_4^{(m)}&=\lambda_6W_2^{(m)},                    \\
  \partial_zW_1^{(m)}&=\lambda_9W_3^{(m)},                    \\
  \partial_yW_4^{(m)}&=\lambda_{10}W_3^{(m)},                 \\
  \partial_zW_2^{(m)}&=\lambda_{13}W_4^{(m)},                 \\
  \partial_yW_3^{(m)}&=\lambda_{14}W_4^{(m)},                 \\
  \partial_z\partial_yW_3^{(m)}&=\lambda_8W_2^{(m)},          \\
  \partial_y\partial_zW_2^{(m)}&=\lambda_{11}W_3^{(m)}.
\end{align}
\end{subequations}
According to the consistency conditions (\ref{lambdaI}), equations (\ref{Schroyz}) describe the self-interactions
of the three 4D fields $\hat B_{\mu\nu}$, $\hat C_{\mu}$ and $\hat D_{\mu}$, and determine their masses.
More specifically, the mass of the 4D KR field $\hat B_{\mu\nu}$ arises from both two extra dimensions,
while the masses of the two 4D vector fields $\hat C_{\mu}$ and $\hat D_{\mu}$ originate from the extra
dimension $z$ and $y$, respectively. There is no self-interaction for the vector field $\hat C_{\mu}$ on
the extra dimension $y$, nor for the vector field $\hat D_{\mu}$ on extra dimension $z$.

For the scalar field $\hat{\zeta}$, the 4D effective action (\ref{4Dactionyz}) reveals that this field is
always massless. It is compatible that there is no equation for the function $W_4^{(m)}$ in eqs. (\ref{Schroyz}).
So, we can take it that the 4D scalar field $\hat{\zeta}$ is localized on the brane.

Likewise, from the consistency conditions (\ref{lambdaI}), eqs. (\ref{1stEqtyz}) describe the interactions
between the four 4D fields $\hat B_{\mu\nu}$, $\hat C_{\mu}$, $\hat D_{\mu}$ and $\hat{\zeta}$, which are 
related to $\lambda_i$. These interactions exist on different extra dimensions, reads:
\begin{itemize}
  \item The interaction between the KR field $\hat B_{\mu\nu}$ and the vector field $\hat C_{\mu}$, and
        the interaction between the vector field $\hat D_{\mu}$ and the scalar field $\hat{\zeta}$ exist
        on the dimension $y$;
  \item The interaction between the KR field $\hat B_{\mu\nu}$ and the vector field $\hat D_{\mu}$, and
        the interaction between the vector field $\hat C_{\mu}$ and the scalar field $\hat{\zeta}$ exist
        on the dimension $z$;
  \item The interaction between the two vector fields $\hat C_{\mu}$ and $\hat D_{\mu}$ exists on both
        two extra dimensions $y$ and $z$.
\end{itemize}

For eqs. (\ref{1stEqtyz}), it can further lead to
\begin{subequations}\label{SchroIntyz}
\begin{align}
  \partial_y^2W_1^{(m)}&=\lambda_3\lambda_5W_1^{(m)},                          \\
  \partial_z^2W_1^{(m)}&=\lambda_4\lambda_9W_1^{(m)},                          \\
  \partial_y^2W_2^{(m)}&=\lambda_3\lambda_5W_2^{(m)},                          \\
  \partial_z^2W_2^{(m)}&=\lambda_6\lambda_{13}W_2^{(m)},                       \\
  \partial_y^2W_3^{(m)}&=\lambda_{10}\lambda_{14}W_3^{(m)},                    \\
  \partial_z^2W_3^{(m)}&=\lambda_4\lambda_9W_3^{(m)},                          \\
  \partial_y^2W_4^{(m)}&=\lambda_{10}\lambda_{14}W_4^{(m)},                    \\
  \partial_z^2W_4^{(m)}&=\lambda_6\lambda_{13}W_4^{(m)},                       \\
  \partial_y^2\partial_z^2W_2^{(m)}&=\lambda_8\lambda_{11}W_2^{(m)},           \\
  \partial_y^2\partial_z^2W_3^{(m)}&=\lambda_8\lambda_{11}W_3^{(m)}.
\end{align}
\end{subequations}
These equations are solvable, and from them we can see that the interaction between the corresponding
two 4D fields brings the same effect on each of the two.

Comparing eqs. (\ref{Schroyz}) with (\ref{SchroIntyz}), we can find
\begin{subequations} \label{lambdaRELAyz}
\begin{align}
  & \lambda_1=\lambda_{12}=\lambda_3\lambda_5=\lambda_{10}\lambda_{14},            \\
  & \lambda_2=\lambda_7=\lambda_4\lambda_9=\lambda_6\lambda_{13},                  \\
  & \lambda_1\lambda_2=\lambda_8\lambda_{11}.
\end{align}
\end{subequations}
These expressions denote the relations of the magnitudes among the self-interactions and the interactions
of the four 4D fields.

Next, we focus on the masses of these 4D fields, and introduce $m_1$, $m_2$, $m_{\text C}$ and $m_{\text D}$
\begin{subequations}  \label{massyz}
\begin{eqnarray}
  &&m_1^2=-\lambda_1,                       \\
  &&m_2^2=-\lambda_2,                       \\
  &&m_{\text C}^2=-\lambda_7,               \\
  &&m_{\text D}^2=-\lambda_{12},
\end{eqnarray}
\end{subequations}
so $m_{\text{KR}}=\sqrt{m_1^2+m_2^2}$ is the mass of the 4D KR field $\hat B_{\mu\nu}$, and $m_{\text C},m_{\text D}$
are the masses of the two 4D vector fields $\hat C_{\mu}$ and $\hat D_{\mu}$, respectively.

By performing further KK decomposition for $W_1^{(m)},W_2^{(m)},W_3^{(m)}$ and $W_4^{(m)}$
\begin{eqnarray} \label{KKdecomposition2yz}
  &&W_1^{(m)}(y,z)=\sum_nw_1^{(m,n)}(y)u_1^{(m,n)}(z),             \\
  &&W_2^{(m)}(y,z)=\sum_nw_2^{(m,n)}(y)u_2^{(m,n)}(z),             \\
  &&W_3^{(m)}(y,z)=\sum_nw_3^{(m,n)}(y)u_3^{(m,n)}(z),             \\
  &&W_4^{(m)}(y,z)=\sum_nw_4^{(m,n)}(y)u_4^{(m,n)}(z),
\end{eqnarray}
eqs. (\ref{Schroyz}) can give rise to
\begin{subequations}  \label{Schroyz1B}
\begin{eqnarray}
  &&-\partial_y^2w_1^{(m,n)}=m_1^2w_1^{(m,n)},                             \\
  &&-\partial_z^2u_1^{(m,n)}=m_2^2u_1^{(m,n)},                             \\
  &&-\partial_z^2u_2^{(m,n)}=m_{\text C}^2u_2^{(m,n)},                     \\
  &&-\partial_y^2w_3^{(m,n)}=m_{\text D}^2w_3^{(m,n)},
\end{eqnarray}
\end{subequations}
where $w_i^{(m,n)}(y)$ and $u_i^{(m,n)}(z)$ $(i=1,2,3,4)$ denote the KK modes of the four 4D fields with respect
to the extra dimensions. From eqs. (\ref{lambdaI}) and eqs. (\ref{massyz}), we can regard $m_{\text{KR}}$,
$m_{\text C}$ and $m_{\text D}$ as representing the masses of the 4D fields $\hat B_{\mu\nu}$, $\hat C_{\mu}$
and $\hat D_{\mu}$, respectively. Then, according to eqs. (\ref{Schroyz1B}), $m_1$, $m_2$, $m_{\text C}$
and $m_{\text D}$ can further be interpreted as the masses of the KK modes of these 4D fields.

Moreover, in eqs. (\ref{Schroyz}) and eqs. (\ref{Schroyz1B}), the two corresponding equations exhibit
the common Schr\"{o}dinger-like form. In particular, their eigenvalues are related through the relations
(\ref{massyz}). Consequently, the two orders $m$ and $n$ are also correlated. For example, in eqs.
(\ref{Schroyz}a) and (\ref{Schroyz1B}a), when the eigenvalues $\lambda_1=m_1=0$, both eigenfunctions
are constants and the orders $m=n=0$.

Then, from eqs. (\ref{Schroyz1B}), it can be known that all the zero mode solutions $w_1^{(0,0)}$, $u_1^{(0,0)}$,
$u_2^{(0,0)}$, and $w_3^{(0,0)}$ are constants, so these zero modes cannot be normalized and the 4D fields
$\hat B_{\mu\nu}$, $\hat C_{\mu}$ and $\hat D_{\mu}$ cannot be localized on the brane. For the KK modes
$w_1^{(m,n)}, u_1^{(m,n)},u_2^{(m,n)}$, and $w_3^{(m,n)}$, each of them forms a series of continuum
spectrum of mass.

In addition, there is no tachyonic modes for the KK modes $w_1^{(m,n)}$, $u_1^{(m,n)}$, $u_2^{(m,n)}$
or $w_3^{(m,n)}$, all of these KK modes are stable.

Therefore, in the case of the extra dimensions $\mathcal{R}_1\times\mathcal{R}_1$, the 4D fields
$\hat B_{\mu\nu}$, $\hat C_{\mu}$ and $\hat D_{\mu}$ cannot be localized on the brane. However, the 4D
scalar field $\hat{\zeta}$ remains massless, and can always be localized on the brane. In the above
analysis, the specific brane model is not required, so the above localization results of these 4D fields
remains unchanged regardless of whether there is a flat or bent brane.

\subsection{The extra dimensions $\mathcal{R}_1\times\mathcal{S}_1$} \label{sec2-2}

For the extra dimensions with manifold $\mathcal{R}_1\times\mathcal{S}_1$, the line element ansatz is
\begin{eqnarray} \label{metricztheta}
  ds^2=a^2(z,\Theta)(\hat g_{\mu\nu}dx^{\mu}dx^{\nu}+dz^2+d\Theta^2),
\end{eqnarray}
where $z$ is the coordinate of the non-compact extra dimension, and $d\Theta=R_0d\theta$ denotes the
compact extra dimension with the radius $R_0$ and $\theta\in[0,2\pi)$.

The compact extra dimension is assumed to have a enough small radius, and its coordinate 
$\Theta\in[0,2\pi R_0)$. In this case, if the various KK modes are smooth along the compact dimension, 
they can be normalized over it. Indeed, this characteristic does not require further assumptions about 
the brane model. 

For this case, we conduct the following KK decomposition
\begin{subequations}\label{decomposition1ztheta}
\begin{eqnarray}
      B_{\mu\nu}&=&\sum_m \hat B_{\mu\nu}^{(m)}(x^{\sigma})W_1^{(m)}(z,\Theta),                   \\
      B_{\mu z} &=&\sum_m \hat C_{\mu}^{(m)}(x^{\sigma})W_2^{(m)}(z,\Theta),                      \\
      B_{\mu \Theta} &=&\sum_m \hat D_{\mu}^{(m)}(x^{\sigma})W_3^{(m)}(z,\Theta),                 \\
      B_{z\Theta}    &=&\sum_m \hat{\zeta}^{(m)}(x^{\sigma})W_4^{(m)}(z,\Theta).
\end{eqnarray}
\end{subequations}
The metric (\ref{metricztheta}) and the above decomposition are similar to those of the case
for the extra dimensions $\mathcal{R}_1\times\mathcal{R}_1$, the difference emerges when introducing the
further decomposition for $W_1^{(m)}(z,\Theta),...,W_4^{(m)}(z,\Theta)$, so with the similar methods, we
can also get
\begin{subequations} \label{Schroytheta}
\begin{eqnarray}
  &&\partial_z^2W_1^{(m)}=\lambda_1W_1^{(m)},                                   \\
  &&\partial_{\Theta}^2W_1^{(m)}=\lambda_2W_1^{(m)},                            \\
  &&\partial_{\Theta}^2W_2^{(m)}=\lambda_7W_2^{(m)},                            \\
  &&\partial_z^2W_3^{(m)}=\lambda_{12}W_3^{(m)}.
\end{eqnarray}
\end{subequations}
Likewise, we introduce $m_1$, $m_2$, $m_{\text C}$ and $m_{\text D}$
\begin{eqnarray} \label{massztheta}
  &&m_1^2=-\lambda_1,                     \\
  &&m_2^2=-\lambda_2,                     \\
  &&m_{\text C}^2=-\lambda_7,             \\
  &&m_{\text D}^2=-\lambda_{12},
\end{eqnarray}
where $m_{\text{KR}}=\sqrt{m_1^2+m_2^2}$ is the mass of the 4D KR field $\hat B_{\mu\nu}$, $m_{\text C}$ and
$m_{\text D}$ are the masses of the two 4D vector fields $\hat C_{\mu}$ and $\hat D_{\mu}$, respectively. By
performing the further decomposition
\begin{eqnarray} \label{KKdecomposition2ztheta}
  &&W_1^{(m)}(z,\Theta)=\sum_nw_1^{(m,n)}(z)e^{il_n\Theta},             \\
  &&W_2^{(m)}(z,\Theta)=\sum_nw_2^{(m,n)}(z)e^{il_n\Theta},             \\
  &&W_3^{(m)}(z,\Theta)=\sum_nw_3^{(m,n)}(z)e^{il_n\Theta},             \\
  &&W_4^{(m)}(z,\Theta)=\sum_nw_4^{(m,n)}(z)e^{il_n\Theta},
\end{eqnarray}
eqs. (\ref{Schroytheta}) can reduce to
\begin{subequations} \label{Schroztheta1B}
\begin{eqnarray}
  -\partial_z^2w_1^{(m,n)}(z)&=&m_1^2w_1^{(m,n)}(z),                             \\
  l_n^2&=&m_2^2,                                                                 \\
  l_n^2&=&m_{\text C}^2,                                                         \\
  -\partial_z^2w_3^{(m,n)}(z)&=&m_{\text D}^2w_3^{(m,n)}(z).
\end{eqnarray}
\end{subequations}
Here, the azimuthal number $l_n=n/R_0$, where $n$ is an integer according to the periodic boundary condition.
A concrete function $e^{il_n\Theta}$ is used to denote the component of the compact extra dimension, as 
the more feasible results (\ref{Schroztheta1B}b,\ref{Schroztheta1B}c) can be obtained.

From these expressions (\ref{Schroztheta1B}), firstly we can find that the mass $m_{\text{KR}}$ of the 4D 
KR field $\hat B_{\mu\nu}$ comes from both the extra dimensions $z$ and $\Theta$, while the masses of the
two vector fields $\hat C_{\mu}$ and $\hat D_{\mu}$ come from the dimension $\Theta$ and the dimension $z$,
respectively.

Then, for eqs. (\ref{Schroztheta1B}a) and (\ref{Schroztheta1B}d), it is clear that their zero mode solutions $w_1^{(0,0)}$
and $w_3^{(0,0)}$ are constants, so these two zero modes cannot be normalized, and the two 4D fields 
$\hat B_{\mu\nu}, \hat D_{\mu}$ cannot be localized on the brane. Concerning the 4D vector field $\hat C_{\mu}$,
it derives its mass only from the compact dimension $\Theta$, so this field can be localized on the brane.
For the 4D scalar field $\hat{\zeta}$, it is massless and can always be localized on the brane. In 
addition, similar to eqs. (\ref{lambdaRELAyz}), there also be 
\begin{eqnarray} \label{massRevztheta}
  &&m_1^2=m_{\text D}^2,                     \\
  &&m_2^2=m_{\text C}^2.
\end{eqnarray}

Therefore, in the case of the extra dimensions $\mathcal{R}_1\times\mathcal{S}_1$, the 4D KR field $\hat B_{\mu\nu}$
and the 4D vector field $\hat D_{\mu}$ cannot be localized on the brane. The 4D vector field $\hat C_{\mu}$ and
the 4D scalar field $\hat{\zeta}$ can be localized on the brane, and the latter field always be massless.

\section{Concrete Braneworld Model}\label{Loc}

In this section, we will discuss the localization of the 6D KR field within a concrete brane model, which
is in the case of the extra dimensions $\mathcal{R}_1\times\mathcal{S}_1$. As discussed in section \ref{method},
the 4D KR field $\hat B_{\mu\nu}$ in this scenario cannot be localized on the brane with minimal coupling.
To address this issue, we will introduce the background scalar coupling below.

In this setup, the line element of the 6D bulk spacetime is assumed as
\begin{eqnarray} \label{metricMin}
  ds^2=a^2(y)\eta_{\mu\nu}dx^{\mu}dx^{\nu}+dy^2+a^2(y)d\Theta^2,
\end{eqnarray}
where $\eta_{\mu\nu}$ is the metric of the flat brane, and the warp factor $a(y)$ is a function of the extra
dimension $y$. Through the coordinate transformation 
\begin{eqnarray} \label{coordinate transformation}
\left\{
  \begin{array}{ll}
    dz&=a^{-1}(y)dy,  \\
    z&=\int a^{-1}(y)dy,
  \end{array}
\right.
\end{eqnarray}
where the boundary condition $z(y=0)=0$, this line element can be expressed in terms of the conformal 
coordinate $z$:  
\begin{eqnarray} \label{metricMinz}
  ds^2=a^2(z)(\eta_{\mu\nu}dx^{\mu}dx^{\nu}+dz^2+d\Theta^2). 
\end{eqnarray}

For this 6D bulk spacetime, a thick brane solution is presented in ref. \cite{WJJjhep2105017}
\begin{subequations} \label{brane model}
\begin{eqnarray}
  \phi(y)&=&v\tanh(ky),                                                        \\
  a(y)&=&e^{-\frac{1}{24}v^2\tanh^2(ky)}\text{sech}^{\frac{v^2}{6}}(ky),
\end{eqnarray}
\end{subequations}
where the background scalar field $\phi(y)$ is a kink function of the extra dimension $y$, $v$ is a dimensionless
parameter and $k$ is a fundamental energy scale with dimension $[k]=L^{-1}$. $1/k$ stands for the thickness
of the brane.

With considering the background scalar coupling, the 6D action for a massless KR field is
\begin{eqnarray} \label{actionF}
  S_{\text{KR}}=-\int d^6x\sqrt{-g}F(\phi)H^{MNL}H_{MNL},
\end{eqnarray}
where factor $F(\phi)$ represents the coupling between the background scalar field and the 6D KR field $B_{MN}$, and it
is a function of the background scalar $\phi$. This coupling function $F(\phi)$ should be nonsingular, and always
be positive to preserve the canonical form of 4D action. Besides, this coupling should return to the minimal
coupling with $F(\phi)=1$ when the background scalar vanishes.

In terms of the metric (\ref{metricMinz}), we perform the KK decomposition (\ref{decomposition1ztheta})
for the 6D KR field $B_{MN}$, and the further decomposition
\begin{eqnarray} \label{KKdecompositionFztheta}
  &&W_1^{(m)}(z,\Theta)=\sum_nw_1^{(m,n)}(z)e^{il_n\Theta}(F(\phi))^{-\frac12},             \\
  &&W_2^{(m)}(z,\Theta)=\sum_nw_2^{(m,n)}(z)e^{il_n\Theta}(F(\phi))^{-\frac12},             \\
  &&W_3^{(m)}(z,\Theta)=\sum_nw_3^{(m,n)}(z)e^{il_n\Theta}(F(\phi))^{-\frac12},             \\
  &&W_4^{(m)}(z,\Theta)=\sum_nw_4^{(m,n)}(z)e^{il_n\Theta}(F(\phi))^{-\frac12}.
\end{eqnarray}
Then the calculation is similar with that given in section \ref{method} and the processes will not be shown
again. Subsequently, with introducing $m_1$, $m_2$, $m_{\text C}$ and $m_{\text D}$ defined as the same as
eqs. (\ref{massztheta}), we can get the following equations
\begin{eqnarray}
  (-\partial_z^2+V(z))w_1^{(m,n)}(z)&=&m_1^2w_1^{(m,n)}(z),             \label{SchrozthetaF1B}      \\
  l_n^2&=&m_2^2,                                                                  \label{SchrozthetaF2B}      \\
  l_n^2&=&m_{\text C}^2,                                                          \label{SchrozthetaF3C}      \\
  (-\partial_z^2+V(z))w_3^{(m,n)}(z)&=&m_{\text D}^2w_3^{(m,n)}(z)              \label{SchrozthetaF4D}
\end{eqnarray}
with the effective potential
\begin{eqnarray} \label{EffPot}
  V(z)=\frac{F''(\phi)}{2F(\phi)}-\frac{F'^2(\phi)}{4F^2(\phi)},
\end{eqnarray}
where the prime denotes the derivative in terms of the coordinate $z$. Likewise, the mass of the 4D KR
field $\hat B_{\mu\nu}$ is $m_{\text{KR}}=\sqrt{m_1^2+m_2^2}$, $m_{\text C}$ represents
the mass for the 4D vector field $\hat C_{\mu}$, and $m_{\text D}$ for the 4D vector field  $\hat D_{\mu}$.

From eqs. (\ref{SchrozthetaF3C}) and (\ref{SchrozthetaF4D}), we can see that the masses $m_{\text C}$
and $m_{\text D}$ arise from the extra dimensions $\Theta$ and $z$, respectively. The 4D vector field
$\hat C_{\mu}$ can be localized on the brane owing to the existence of the compact extra dimension, and the
4D scalar field $\hat{\zeta}$ can also be localized. Besides, from eqs. (\ref{SchrozthetaF2B})$\sim$(\ref{SchrozthetaF4D}),
the masses of different KK modes exhibit the following relations: 
\begin{eqnarray} \label{massRevztheta1}
  m_1^2&=&m_{\text D}^2,                     \\
  m_2^2&=&m_{\text C}^2,                     \\
  m_{\text{KR}}^2&=&\sqrt{m_1^2+m_2^2}=\sqrt{m_{\text D}^2+m_{\text C}^2}. 
\end{eqnarray}
Along the same one extra dimension, the mass of the KK modes for the 4D KR field is equal to the mass of
the vector KK mode of the same order. 

In brane models, it is usually assumed that the extra dimension possesses $\mathbb{Z}_2$ symmetry, so the
effective potential (\ref{EffPot}) should be symmetric concerning the extra dimension $z$. We can see that
this condition is satisfied when the coupling function $F(\phi)$ is of even parity. In the light of the
aforementioned discussions for this  function, its form can be considered as \cite{ExtraDimensionLiu}
\begin{eqnarray} \label{Fphi}
  F(\phi)=\left(1-\frac{\phi^2}{v^2}\right)^t,
\end{eqnarray}
where $t$ is a positive coupling parameter.

Furthermore, since the two Schr\"{o}dinger-like equations (\ref{SchrozthetaF1B}) and (\ref{SchrozthetaF4D})
share the same expression of the effective potential (\ref{EffPot}), we will only discuss the former one
equation for convenience, and it can further be factorized as
\begin{eqnarray} \label{SchrozthetaFFac1B}
  \left(\partial_z+\Gamma'(z)\right)\left(-\partial_z+\Gamma'(z)\right)w_1^{(m,n)}(z)&=&
                                            m_1^2w_1^{(m,n)}(z)
\end{eqnarray}
with
\begin{eqnarray} \label{Gammaprime}
  \Gamma'(z)=\frac{F'(\phi)}{2F(\phi)}.
\end{eqnarray}
This equation can give rise to the zero mode solution
\begin{eqnarray} \label{ZM}
  w_1^{(0,0)}(z)=N_1F(\phi)^{\frac12},
\end{eqnarray}
where $N_1$ is the normalized constant. From the equation (\ref{SchrozthetaFFac1B}), we can see that
the eigenvalues are nonnegative, so there is no tachyonic modes. In view of eqs. (\ref{SchrozthetaF2B})
and (\ref{SchrozthetaF3C}), we can find that all the various KK modes are stable.

\begin{figure} 
\begin{center}
\subfigure[]{\label{FigResVeca}
\includegraphics[width= 0.48\textwidth]{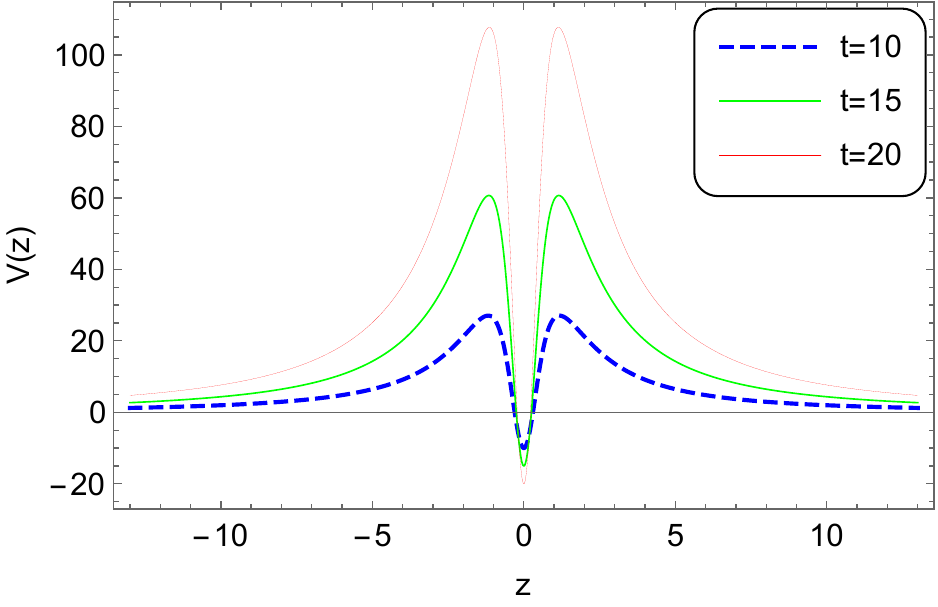}}
\subfigure[]{\label{FigResVecb}
\includegraphics[width= 0.48\textwidth]{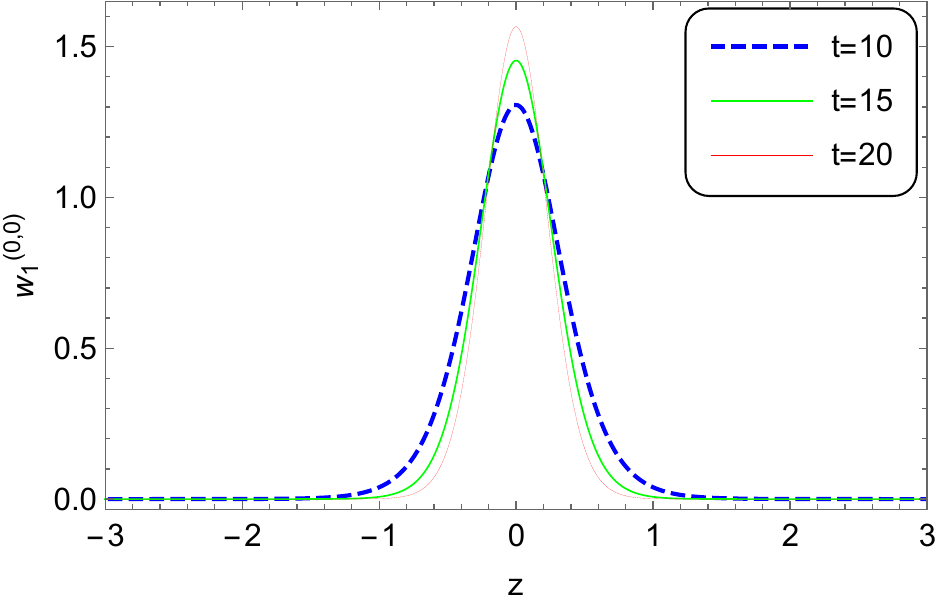}}
\end{center}\vskip -2mm
\caption{The effective potential $V(z)$ and the zero mode $w_1^{(0,0)}(z)$ with parameters
         $v=2$, $k=1$ and $t=10,15,20$.}
 \label{FigResVec}
\end{figure}

Based on the brane model (\ref{brane model}) and the function $F(\phi)$ (\ref{Fphi}), this zero mode (\ref{ZM})
can be expressed as
\begin{eqnarray} \label{ZMy}
  w_1^{(0,0)}(z(y))=N_1\text{sech}^t(ky).
\end{eqnarray}
The normalization of this zero mode requires
\begin{eqnarray} \label{NormCondZM}
  \int|w_1^{(0,0)}(z)|^2dz&=&\int\big|w_1^{(0,0)}(z(y))\big|^2a^{-1}dy     \nonumber     \\
        &=&N_1^2\int e^{\frac{v^2}{24}\tanh^2(ky)}(\text{sech}(ky))^{2t-\frac{v^2}{6}}dy=1.
\end{eqnarray}
It is easy to see that if the coupling parameter $t>v^2/12$, this normalization can be realized, and
both the 4D KR field $\hat B_{\mu\nu}$ and the 4D vector field $\hat D_{\mu}$ can be localized on the
brane. The zero mode $w_1^{(0,0)}(z)$ (\ref{ZM}) is illustrated in figure \ref{FigResVecb} via numerical
methods, with parameters $v=2,k=1$ and $t=10,15,20$. Then, all of the four 4D fields can be localized
on the brane at the same time.

Concerning the effective potential (\ref{EffPot}), in terms of the brane model (\ref{brane model}), there
is no analytical expression for this potential concerning coordinate $z$. Hence, with the coordinate
transformation (\ref{coordinate transformation}), we reexpress this effective potential as
\begin{eqnarray} \label{EffPoty}
  V(z(y))=\frac{a^2F''(\phi)+aa'F'(\phi)}{2F(\phi)}-\frac{a^2F'^2(\phi)}{4F^2(\phi)},
\end{eqnarray}
where the prime denotes the derivative with respect to coordinate $y$. Based on eqs. (\ref{brane model})
and (\ref{Fphi}), there further be
\begin{eqnarray} \label{EffPotzy}
  V(z(y))&=&\frac{1}{12}k^2t\ e^{-\frac{v^2}{12}\tanh^2(ky)}(\text{sech}(ky))^{2+\frac{v^2}{3}}    \nonumber    \\
         & &\times\left[-6(t+2)+(v^2+6t)\cosh(2ky)-v^2\text{sech}^2(ky)\right].
\end{eqnarray}
From this expression, we can obtain the following behaviors for this potential $V(z(y))$:
\begin{eqnarray}
  V(z(y=0))\hspace{0.1cm} &=&  -k^2t,                               \label{EffPotOri}   \\
  V(z(y\rightarrow\pm\infty))   &\rightarrow&  0^+.             \label{EffPotInfit}
\end{eqnarray}
It can be seen that the effective potential (\ref{EffPotzy}) is in a volcano shape. With certain 
values of parameters, we plot this effective potential in figure \ref{FigResVeca} concerning coordinate $z$ using
numerical methods. From this figure \ref{FigResVeca}, we can get that the corresponding massive KK modes
cannot be localized.

On the other hand, it is found that if the coupling parameter $t$ increases, the potential well at the origin
becomes deeper. So, the resonant modes could exist at this time. Referring to the method proposed in
refs. \cite{PRDLYX0904.1785,PRDLYX0907.0910}, the relative probability function for a resonant mode on the
thick brane can be defined as
\begin{eqnarray} \label{ResoP}
  P(m^2)=\frac{\int^{z_b}_{-z_b}|w_1^{(m,n)}(z)|^2dz}{\int^{z_{max}}_{-z_{max}}|w_1^{(m,n)}(z)|^2dz},
\end{eqnarray}
where $2z_b$ denotes the width of the brane, and $z_{max}$ is set as $10z_b$. For the KK modes with enough
larger $m_1^2$ values than the maximum of the corresponding potential, they tend toward plane waves, and
their probabilities approaching $0.1$. The lifetime $\tau$ of a resonance is $\tau\sim\Gamma^{-1}$, where
$\Gamma=\delta m_1$ is the full width at half maximum of the resonant peak.

\begin{figure}[htb]
\begin{center}
\subfigure[$t=10$.]{\label{FigResSpecKR}
\includegraphics[width= 0.38\textwidth]{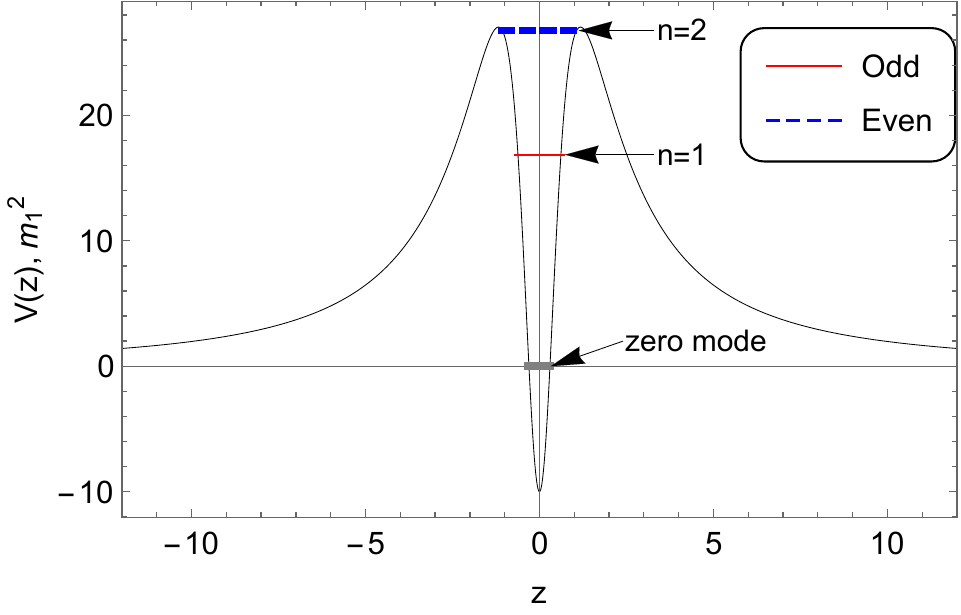}}
\hspace{0.5cm}
\subfigure[$t=10$.]{\label{FigResPKR}
\includegraphics[width= 0.38\textwidth]{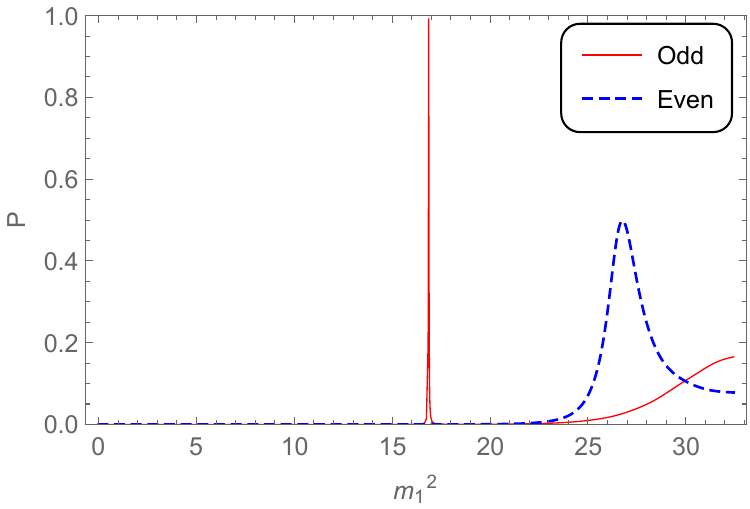}}
\subfigure[$t=15$.]{\label{2FigResSpecKR}
\includegraphics[width= 0.38\textwidth]{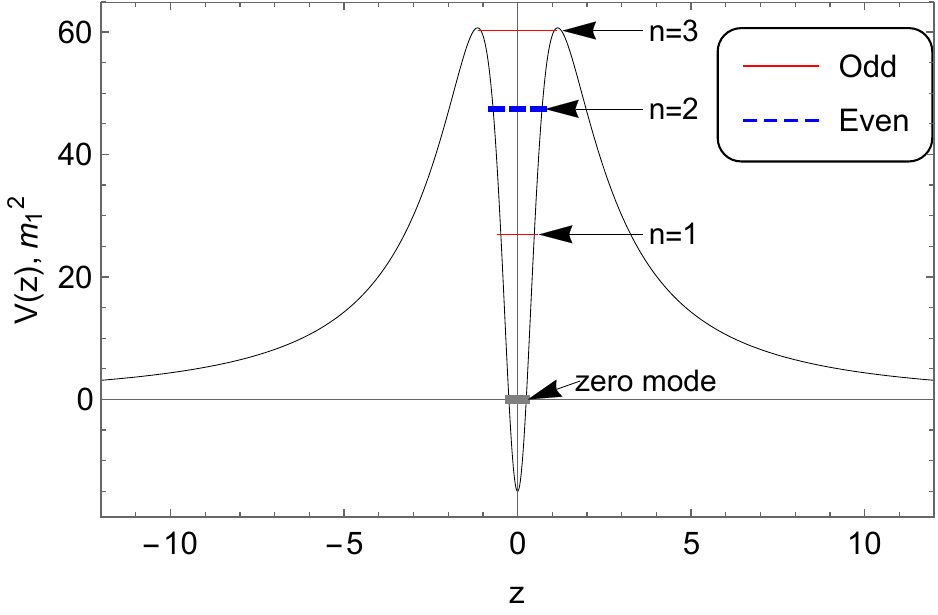}}
\hspace{0.5cm}
\subfigure[$t=15$.]{\label{2FigResPKR}
\includegraphics[width= 0.38\textwidth]{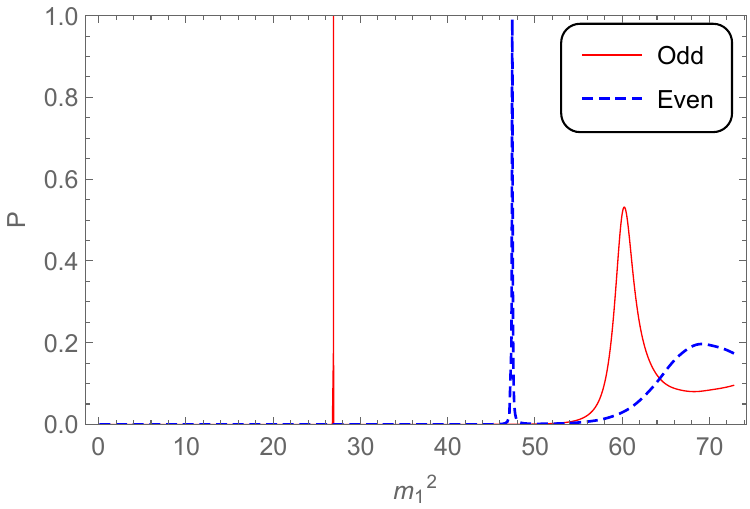}}
\subfigure[$t=20$.]{\label{3FigResSpecKR}
\includegraphics[width= 0.38\textwidth]{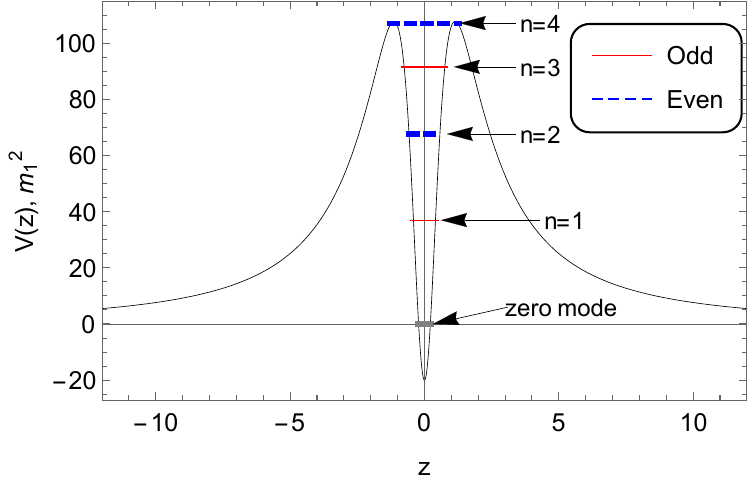}}
\hspace{0.5cm}
\subfigure[$t=20$.]{\label{3FigResPKR}
\includegraphics[width= 0.38\textwidth]{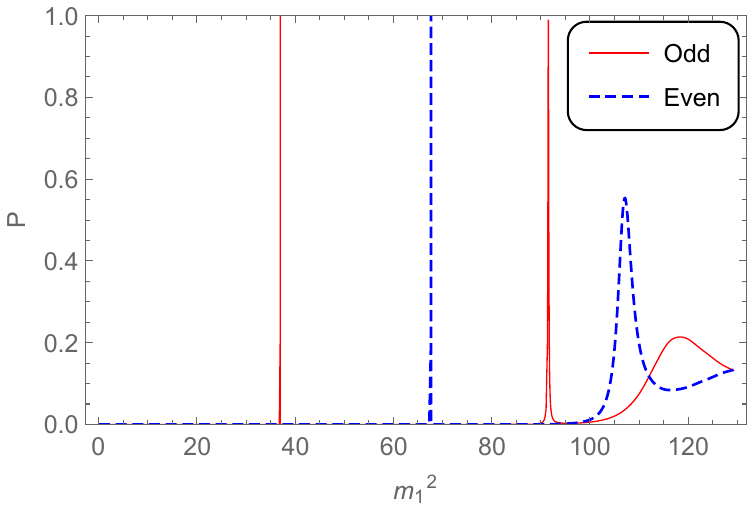}}
\end{center}\vskip -5mm
\caption{The mass spectra, the effective potential $V(z)$, and corresponding relative probability
         $P$ with the coupling parameter $t=10,15,20$. The potential $V(z)$ for the black
         line, the zero mode for the grey line, the even parity resonant KK modes for the blue
         lines, and the odd parity resonant KK modes for the red lines. The parameters are set
         as $v=2$ and $k=1$.}
 \label{FigResSpec-PKR}
\end{figure}

\begin{figure}[htb]
\begin{center}
\subfigure[$w_1^{(0,0)}$.]{\label{figZM}
\includegraphics[width = 0.34\textwidth]{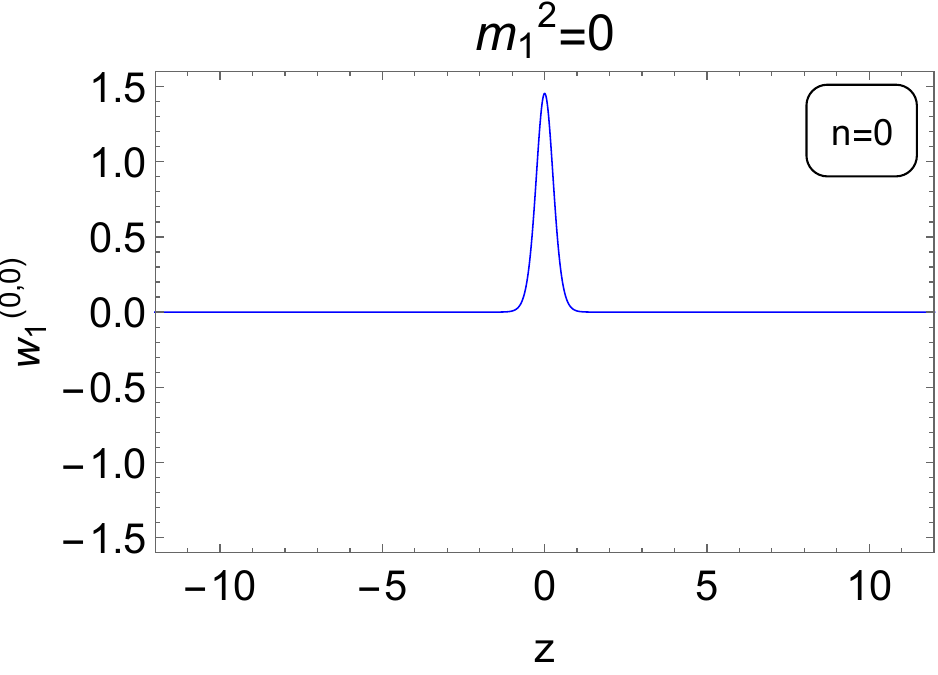}}
\subfigure[$w_1^{(1,1)}$.]{\label{figOdd1}
\includegraphics[width = 0.34\textwidth]{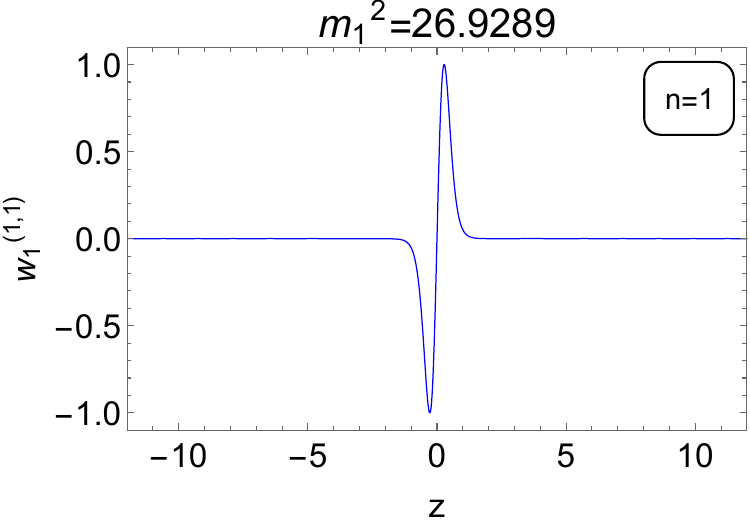}}
\subfigure[$w_1^{(2,2)}$.]{\label{figEven1}
\includegraphics[width = 0.34\textwidth]{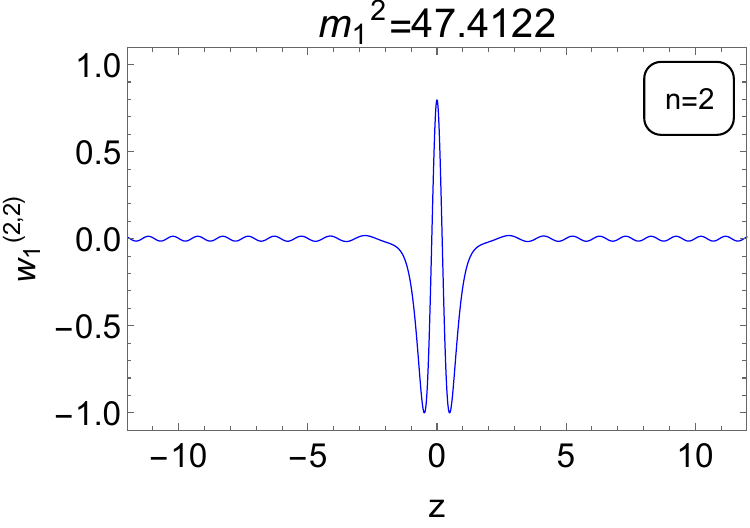}}
\subfigure[$w_1^{(3,3)}$.]{\label{figOdd2}
\includegraphics[width = 0.34\textwidth]{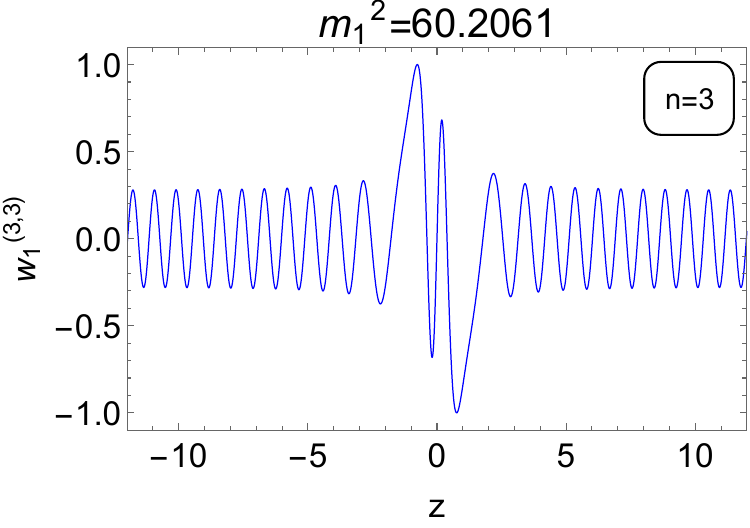}}
\end{center}\vskip -2mm
\caption{The shapes of the zero mode and the resonance modes with different $m_1^{2}$.
         The parameters are set as $v=2,k=1$, and $t=15$.}
 \label{FigResZMKR}
\end{figure}

With the method mentioned above, the resonant modes $w^{(m,n)}_1(z)$ corresponding to the effective potentials
are numerically solved, and plotted in figure \ref{FigResVeca}. The profiles of the relative probabilities $P$
corresponding to different values of coupling parameter are presented in figure \ref{FigResSpec-PKR}. In
these figures, each peak represents a resonant state, and the mass spectra alongside with the effective
potential are also shown in the same figure. In the mass spectra, the ground state is zero mode (bound
state), and all the massive modes are resonant modes. The mass, width, and lifetime of these resonant
modes are detailedly showcased in table \ref{tableRMVolKR}. We can see that the number of the resonances
increases with the coupling parameter $t$. All the resonant modes, as well as the zero mode concerning the
coupling parameter $t=15$ are depicted in figure. \ref{FigResZMKR}, as an illustration.
\begin{table}[tbp]
\centering
\begin{tabular}{|c|c|c|c|c|c|c|}
    \hline
    $t$                 & $V^{\text{max}}$      & $n$         & $m_1^2$ &
    $m_1$               & $\Gamma$              & $\tau$
    \\
    \hline
    $10$                & $27.0386$             & $1$         & $16.8635$ &
    $4.1065$            & $3.516\times10^{-3}$  & $284.4243$
    \\
                        &                       & $2$         & $26.7626$ &
    $5.1733$            & $0.1997$              & $5.0082$
    \\
    \hline
    $15$                & $60.6563$             & $1$         & $26.9289$ &
    $5.1893$            & $5.127\times10^{-6}$  & $1.951\times10^5$
    \\
                        &                       & $2$         & $47.4122$ &
    $6.8857$            & $7.385\times10^{-3}$  & $135.4123$
    \\
                        &                       & $3$         & $60.2461$ &
    $7.7618$            & $0.1823$              & $5.4856$
    \\
    \hline
    $20$                & $107.7140$            & $1$         & $36.9517$ &
    $6.0788$            & $1.729\times10^{-9}$  & $5.785\times10^{8}$
    \\
                        &                       & $2$         & $67.6505$ &
    $8.2250$            & $4.978\times10^{-5}$  & $2.009\times10^4$
    \\
                        &                       & $3$         & $91.5361$ &
    $9.5675$            & $0.0101$              & $99.1483$
    \\
                        &                       & $4$         & $107.1102$ &
    $10.3494$           & $0.1725$              & $5.7979$
    \\
    \hline
\end{tabular}
\caption{The mass, width, and lifetime of resonant KK modes. The parameters
        are set as $k=2$ and $v=1$. }
    \label{tableRMVolKR}
\end{table}

Therefore, by introducing the background scalar coupling, the entire set of 4D fields-the KR field,
the two vector fields, and the scalar field-can be localized on the thick Minkowski brane under 
the condition $t>v^2/12$. In addition, the 4D scalar field always be massless. It is further found 
that for the 4D KR field and the one 4D vector field whose mass originates from the non-compact 
extra dimension, the resonant KK modes could emerge near the origin of this extra dimension. The 
number of the resonant KK modes increases with the coupling parameter $t$.

\section{Conclusions}\label{Cons}

In this paper, we investigate the localization of the KR field in the 6D bulk spacetime with codimension-2.
By adopting a general KK decomposition, the 6D KR field is expanded into one 4D KR field,
two 4D vector fields and one 4D scalar field. We then perform a detailed localization analysis for
these four components under two different configurations of the extra-dimensional manifold:
$\mathcal{R}_1\times\mathcal{R}_1$ and $\mathcal{R}_1\times\mathcal{S}_1$. Finally, based on a concrete
6D brane model, we explore the localization behavior of the 6D KR field with consideration of the background
scalar coupling.

In the case of the extra dimensions $\mathcal{R}_1\times\mathcal{R}_1$, neither the 4D KR field nor the
two 4D vector fields can be localized on the brane. However, the 4D scalar field can be localized on the brane,
and it remains massless. In contrast, for the case of the extra dimensions
$\mathcal{R}_1\times\mathcal{S}_1$, one of the 4D vector fields, along with the 4D scalar field, can be localized
on the brane, while the 4D KR field and the other one 4D vector field remain non-localizable. Despite these
differences, both cases share several common features, reads
\begin{itemize}
  \item The mass of the 4D KR field arises from both extra dimensions, while the masses of the two 4D vector fields
        originate separately from different one of the two extra dimensions;
  \item {Along the same one extra dimension, the mass of the KK modes for the 4D KR field is equal to the mass of 
        the vector KK mode of the same order;} 
  \item The 4D scalar field remains massless, and can always be localized.
\end{itemize}
In addition, the interactions exist between these 4D fields, except for the pair consisting of the 4D KR
field and the 4D scalar field, for which no direct interaction occurs.

Finally, with a concrete 6D brane model, which is in the case of extra dimensions $\mathcal{R}_1\times\mathcal{S}_1$,
the localization analysis is conducted. In this setup, the background scalar coupling is introduced, and a
coupling factor exists in the action of the 6D KR field. It is demonstrated that, in addition to the previously
localized one 4D vector field and 4D scalar field, the 4D KR field and the other one 4D vector field can also be
localized on the brane under the condition of coupling parameter $t>v^2/12$. For these two 4D fields, the resonant KK
modes are found to emerge near the origin of the non-compact extra dimension, and their numbers increase
with the coupling parameter $t$. In the meantime, compared with the minimal coupling case, introducing the
background scalar coupling does not alter the mass relations among the KK modes of these four 4D fields.

\acknowledgments

This work is supported by the National Natural Science Foundation of China (Grants No. 11305119), the Natural
Science Basic Research Plan in Shaanxi Province of China (Program No. 2020JM-198), the Fundamental Research
Funds for the Central Universities (Grants No. JB170502), and the 111 Project (B17035).

\end{document}